# Large-area polycrystalline α-MoO3 thin films for IR photonics

**Maria Cristina Larciprete[1], Daniele Ceneda[1], Chiyu Yang[2], Sina Abedini Dereshgi[3], Federico Vittorio Lupo[4], Maria Pia Casaletto[5], Roberto Macaluso[4], Mauro Antezza[6,7], Zhuomin M. Zhang[2], Marco Centini [1] and Koray Aydin [3]**

[1] Dipartimento di Scienze di Base ed Applicate per l'Ingegneria, Sapienza Università di Roma, Rome, 00161 Italy
[2] George W. Woodruff School of Mechanical Engineering, Georgia Institute of Technology, Atlanta, GA 30332, USA
[3] Department of Electrical and Computer Engineering, Northwestern University, Evanston, Illinois 60208, USA
[4] Department of Engineering, University of Palermo, Palermo, 90128, Italy
[5] National Research Council (CNR), Institute for Nanostructured Materials (ISMN), Palermo, 90146, Italy
[6] Laboratoire Charles Coulomb (L2C), UMR 5221 CNRS-Université de Montpellier, F- 34095 Montpellier, France
[7] Institut Universitaire de France, 1 rue Descartes, F-75231 Paris Cedex 05, France

E-mail: marco.centini@uniroma1.it

## Abstract

In recent years, excitation of surface phonon polaritons (SPhPs) in van der Waals materials received wide attention from the nanophotonics community. Alpha-phase Molybdenum trioxide ($\alpha$-$MoO_3$), a naturally occurring biaxial hyperbolic crystal, emerged as a promising polaritonic material due to its ability to support SPhPs for three orthogonal directions at different wavelength bands (range 10–20 μm). Here, we report on the fabrication and IR characterization of large-area (over 1 $cm^2$ size) $\alpha$-$MoO_3$ polycrystalline films deposited on fused silica substrates by pulsed laser deposition. Single $\alpha$-phase $MoO_3$ films exhibiting a polarization-dependent reflection peak at 1006 $cm^{-1}$ with a resonance Q-factor as high as 53 were achieved. Reflection can be tuned via changing incident polarization with a dynamic range of $\Delta R=0.3$ at 45° incidence angle. We also report a polarization-independent almost perfect absorption condition ($R<0.01$) at 972 $cm^{-1}$ which is preserved for a broad angle of incidence. The development of a low-cost polaritonic platform with high-Q resonances in the mid-infrared (mid-IR) range is crucial for a wide number of functionalities including sensors, filters, thermal emitters, and label-free biochemical sensing devices. In this framework our findings appear extremely promising for the further development of lithography-free, scalable films, for efficient and large-scale devices operating in the free space, using far-field detection setups.

## 1. Introduction

Advances in nanophotonics have enabled the miniaturization of optical components due to the exploitation of surface plasmon polaritons (SPPs) [1] in the visible range that can strongly localize electromagnetic fields to small volumes. Recently, doped semiconductors [2,3] and graphene [4] have been proposed to extend SPPs to the 2–8 μm range. Also, nanoantennas have been employed to achieve electric field localization in the mid-IR range for sensing applications.

However, in order to take full advantage of surface-enhanced infrared absorption (SEIRA) techniques, a precise positioning of the analyte is required [5].

Moving toward the 8–20 μm wavelength range, where vibrational absorption peaks provide relevant information on molecular bonds, the SPP approach is less effective due to the poor field confinement at longer wavelengths [6]. Moreover, for the development of a complete IR photonic platform, miniaturization, and integration of optical components with the chip-scale platforms using facile fabrication techniques [7,8] is highly desired. A conventional IR polarizer is nowadays designed using state-of-the-art holographic techniques [9] onto IR transparent support ($CaF_2$, ZnSe, $BaF_2$) with typical transmission losses of about 30%. Furthermore, the surface of such holographic grid polarizers is extremely delicate, and touching is absolutely to be avoided. Similarly, optical components such as polarization rotators have been realized using artificial metasurfaces [10] in the wavelength range up to 10 μm. Functionality at longer wavelengths is achieved using a combination of two parallel polarizers and by tilting the plates with respect to each other. Given the complexity of these components, their integration with the chip-scale platform can, therefore, be prohibitive and the challenge for an efficient, integrated and robust IR photonic platform persists.

Recent promising solutions are based on the exploitation of polar materials [11] including ultra-thin van der Waals (vdW) materials such as $MoO_3$, $MoS_2$, $Ga_2O_3$, hBN [12,13]. Besides their strong anisotropy related to optical phonons (ideal for polarization rotation and control), they allow strong field localization by the excitation of surface waves called surface phonon polaritons (SPhPs), achieved through the coupling of the electromagnetic field with lattice vibrations. Several works reported on the great potential of polar materials for mid-IR sensing applications up to the terahertz (THz) regime [14,15] and for the realization of compact IR photonic devices [16,17].

Among vdW materials, Molybdenum trioxide ($\alpha$-$MoO_3$) is attracting a great deal of attention [18] as it supports SPhPs in three different wavelength bands for the three orthogonal directions (range 10-20 μm), rendering this material a naturally hyperbolic and biaxial material [19,20]. Increased versatility can be obtained by combining it with other materials. Recent results show that $\alpha$-$MoO_3$ can be combined with vanadium dioxide, ($VO_2$, a phase change material that undergoes insulator to metal phase transition at a temperature of 68° C) [21,22] in order to dynamically tune the polariton resonances. A metamaterial approach has also been proposed based on the random nanostructuring of $\alpha$-$MoO_3$ with subwavelength dielectric elements (i.e., air ellipsoidal inclusions). This scheme could increase design versatility as well as tuning and hybridization of polariton modes [23].

Despite huge potential of this promising material, the development of a novel, highly versatile and compact $\alpha$-$MoO_3$-based IR photonics platforms is hampered by the lack of availability of high-quality scalable films and/or multilayer stacks. $\alpha$-$MoO_3$ for IR photonics and polaritonics is mostly used in the form of physical vapor deposition (PVD)-grown crystalline flakes. Although flakes allow exciting results in terms of hyperbolic phonon polariton excitation along x- and y-directions, there are several drawbacks that might limit the wide adaption of flakes geometries: the existing alignment techniques for flakes with a few tens of nanometers thickness are challenging; the flakes often have irregular shape preventing a good propagation of SPhPs; the dimensions of the flakes are usually limited to few hundreds of μm at most, therefore the large area or integrated/multifunctional devices are not practical. The fabrication process for obtaining such flakes is very complex, requiring investigation of strategies to create efficient conditions for films growth such as high temperatures (e.g., 780°C [24]). Furthermore, a successive mechanical exfoliation process for transferring the desired $MoO_3$ 2D film onto a substrate of interest is needed [25]. In reference [26] a high confinement of near field signal corresponding to a Q factor of 40 has been reported in $\alpha$-$MoO_3$ covering submicron-width trenches. These flakes are, however, difficult to handle and integrate in a practical device while keeping low fabrication costs. Moreover, they are relatively small for far-field applications since their dimensions often reach the diffraction limit of 10–20 μm range IR radiation thus requiring expensive and state of the art near-field detecting schemes.

The realization of single $\alpha$–phase, oriented, large area $MoO_3$ film is still an open technological challenge. Atomic layer deposition (ALD), has been used to obtain good quality $\alpha$-phase $MoO_3$ films, but only after 500 °C post-growth annealing [27]. This was necessary because deposition temperatures higher than 200 °C interfered with the stability of the employed precursor. Furthermore, a long annealing time (> 1h) was required when the annealing was performed at lower temperatures than 500°C. This means that, according to [28], ALD cannot be used to perform $\alpha$-phase $MoO_3$ films deposition in one single step, making more difficult a possible integration of the $MoO_3$ film within a multilayer structure. ALD is, furthermore, an expensive tool employing hazardous precursors gases and even higher temperatures are needed for depositing $MoO_3$ by sublimation [29].

Conventional sputtering techniques were also employed to deposit $MoO_3$ films at room temperature. This led to a multi-phase crystalline $MoO_3$ film only after a post-growth annealing process [28-30]. When the annealing is performed at temperatures greater than 400 °C, it produces monoclinic $\beta$–phase $MoO_3$ films, not useful to exploit an optical phonons response [31]. Pulsed laser deposition (PLD) is a versatile and low-cost deposition technique which has already been employed for the deposition of α-phase $MoO_3$ films at 500 °C [32] and other metal oxides such as $VO_2$ [33], and ZnO [34]. Compared with the exfoliation technique, it allows depositing large area $MoO_3$ films, which can be much more easily handled and integrated into a multilayer structure. However, to the best of our knowledge, a detailed IR characterization aimed at the identification of possible applications of the obtained films has not been reported so far for large area $MoO_3$ films deposited either by PLD or ALD. In the following, we show that PLD can be employed to obtain α-$MoO_3$ films at lower temperatures (e.g. 400 °C), without using harmful precursor gases normally employed by ALD and without the need of any post-growth annealing. Optical IR reflection spectra reveal a remarkable enhanced tunability of the reflection peak related to the z-axis phonon response as a function of the incident electric field polarization. Moreover, a polarization-independent perfect absorption condition is achieved for a broad angle of incidence. These features are not displayed from a single crystal flake. Our results show, for the first time, interesting possibilities for large-scale, lithography-free, polycrystalline $MoO_3$ film to be employed for IR signal management.

## 2. **Sample Fabrication**

### *2.1 $MoO_3$ Deposition.*

The mainly investigated structure in this study is composed of a 2200 nm (average thickness) $MoO_3$ film deposited on a fused silica substrate (Figure 1a) using pulsed laser deposition at 400°C and 0.1 mbar of oxygen pressure. The PLD system employed uses a Q-switched tripled Nd:YAG laser (Quantel mod. YG78C20, λ = 355 nm) generating 6 ns width pulses with an energy of 80 mJ per pulse [32, 33, 35]. The density of energy was maintained at 1.2 J $cm^{-2}$, and the repetition rate was 4 Hz. The $MoO_3$ target was a 1-inch diameter, 0.25-inch-thick disk (purity 99.9%).

Before each deposition, the substrates were cleaned in an ultrasonic bath with acetone, subsequently rinsed with isopropanol and then dried with compressed air. After cleaning, each substrate was clamped onto an electrical heater, which allows achieving temperatures as high as 800 °C. The heater was then placed inside a vacuum bell jar where oxygen gas can be introduced through an electromechanical valve to maintain the desired pressure.

The PLD deposition yields better crystallinity than sputtering due to its higher kinetic energy of ablated species. Furthermore, the PLD setup allows extremely versatile deposition conditions. Thus, the proper choice of deposition parameters and the resulting fabrication constraints is of crucial importance. In the present work, we focus attention on the best choice of parameters for the narrow band IR polarization filter functionality.

### *2.2 Structural and Morphological Characterization.*

X-ray diffraction (XRD) measurements were performed at room temperature to evaluate the crystalline structure of the deposited layers. XRD analysis was performed by using a D5005 diffractometer (Bruker AXS, Karlsruhe, Germany) equipped with a Cu Kα (1.5406 Å) source and operating at 40 kV and 30 mA. The following experimental conditions were used: 5s acquisition time, 0.05° step in a 5° - 90° 2Θ angular range. XRD patterns showed that the high-quality $MoO_3$ films deposited at 400°C exhibited the stable orthorhombic α-phase of $MoO_3$, as shown in Figure 1(b). For the sake of completeness, we include in the supporting material the XRD pattern for a similar sample, deposited at lower temperature (i.e., 200 °C), showing a monoclinic-only phase of the $MoO_3$ film (Figure S1).

The surface morphology of the $MoO_3$ thin films has been characterized by using an Anfatech high speed atomic force microscope (AFM). Consistently with X-ray diffraction measurements, AFM images of surface morphology shown in Figures 1c and 1d revealed a grain distribution in the thin film. The average grain size is around 400 nm and root-mean-square (RMS) roughness is about 100 nm. After the deposition, the film thickness was assessed by profilometry using a Dektak 150 profilometer. The average thickness was found to be approximately 2200 nm. Details on the profilometer measurements and a picture of the Sample are reported in Figure S2 of the supporting material file.

## 3. **Results and Discussion**

### *3.1 Polarization-dependent reflection measurements.*

IR reflection measurements have been performed using a FT-IR interferometer (Invenio-R, Bruker) in the spectral range 6000-400 $cm^{-1}$. The IR source was a glow-bar while the detector is based on deuterated triglycine sulfate (DTGS) pyroelectric detector.

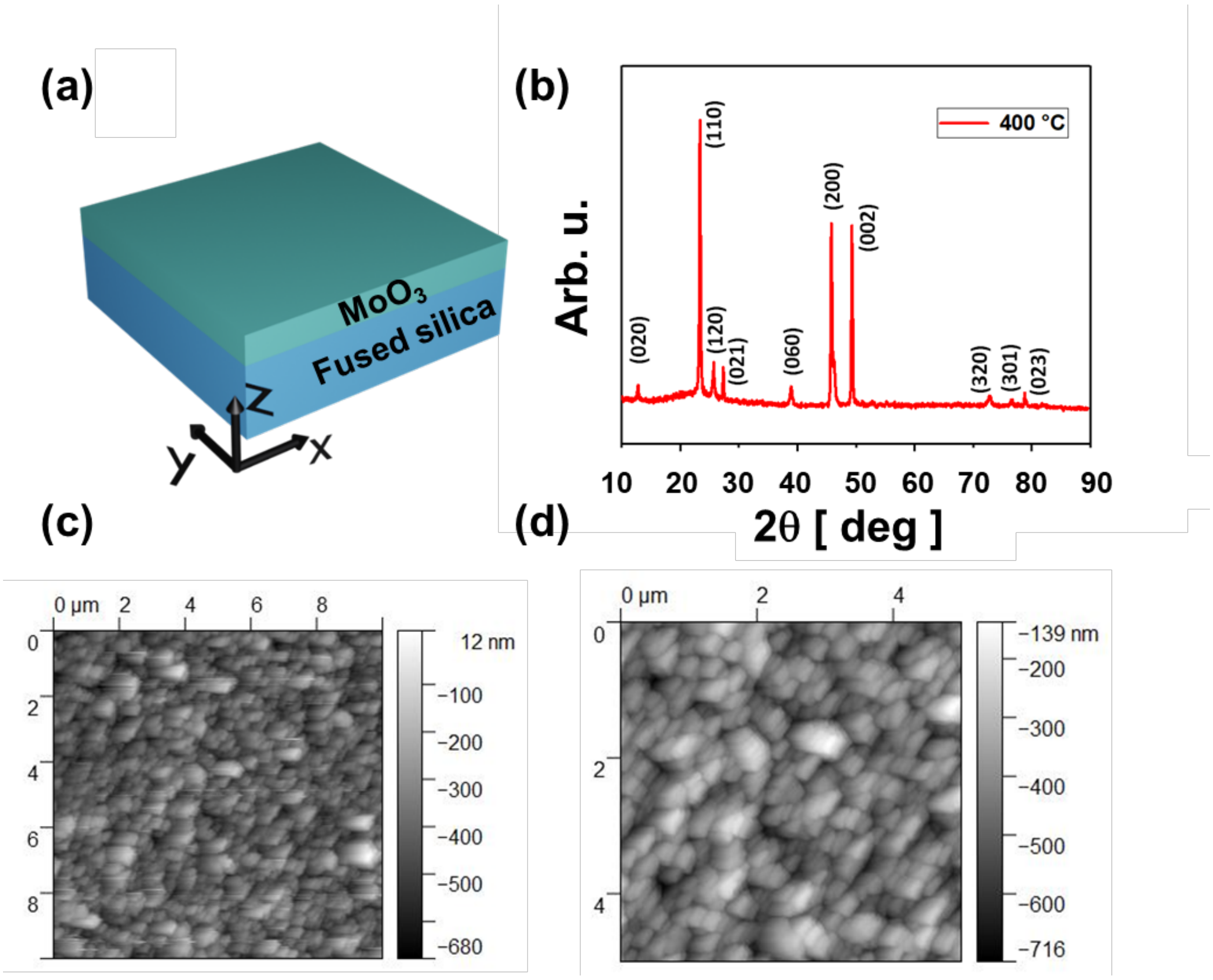

**Figure 1.** (a) Sketch of investigated sample. (b) X-Ray diffraction (XRD) pattern of a $MoO_3$ film deposited onto fused silica by PLD at 400 °C and 0.1 mbar oxygen pressure. The assigned peaks correspond to the orthorhombic phase of $MoO_3$ (ICDD 01-078-4612 card). (c-d) AFM images of a $MoO_3$ film deposited onto fused silica by PLD: (c) image area 10 × 10 μm$^2$; (d) image area 5× 5 μm$^2$.

A total of 64 interferograms were acquired for each measurement, with a spectral resolution of 1 cm$^{-1}$. A sample area of 3x3 mm$^2$ was selected during IR data acquisition using knife-edge apertures. The FT-IR platform is equipped with a reflectance unit allowing to set the angles of incidence and reflectance, from almost normal incidence (about 13°) to grazing angles (85°) as illustrated in Figure (2a). The polarization state of incident light was selected using a holographic polarizing filter with a motorized mounter. Two different sets of measurements were performed with incidence angles of 15° and 45°, respectively. Specifically, the reflectance spectra were recorded as a function of different linear polarization states of the incoming light. The measured spectral reflectance curves for different incidence angles and polarization of the incoming beam are shown in Figure (2b) and Figure (2c) for 15° and 45° incidence angles, respectively. Here 0° polarization angle stands for p-polarized light while 90° stands for s-polarized light.

From Figure (2b) we note that the polycrystalline nature of the laser-deposited $MoO_3$ simultaneously unveils, also at quasi normal incidence, the three Reststrahlen bands associated to alpha phase of $MoO_3$: the x-Reststrahlen band, corresponding to the frequency range from 820 cm$^{-1}$ to 972 cm$^{-1}$; the *y*-Reststrahlen band, extending at lower frequencies, between 545 cm$^{-1}$ and 851 cm$^{-1}$ and the *z*-Reststrahlen band, which is located between 962 cm$^{-1}$ and 1010 cm$^{-1}$ which is also partially overlapped with the Reststrahlen band of glass substrate (fused silica) at 1000 cm$^{-1}$ and 1300 cm$^{-1}$ [18]. Moreover, the polarization-resolved set of measurements, shows that at quasi-normal incidence, the sample exhibits negligible in-plane anisotropy.

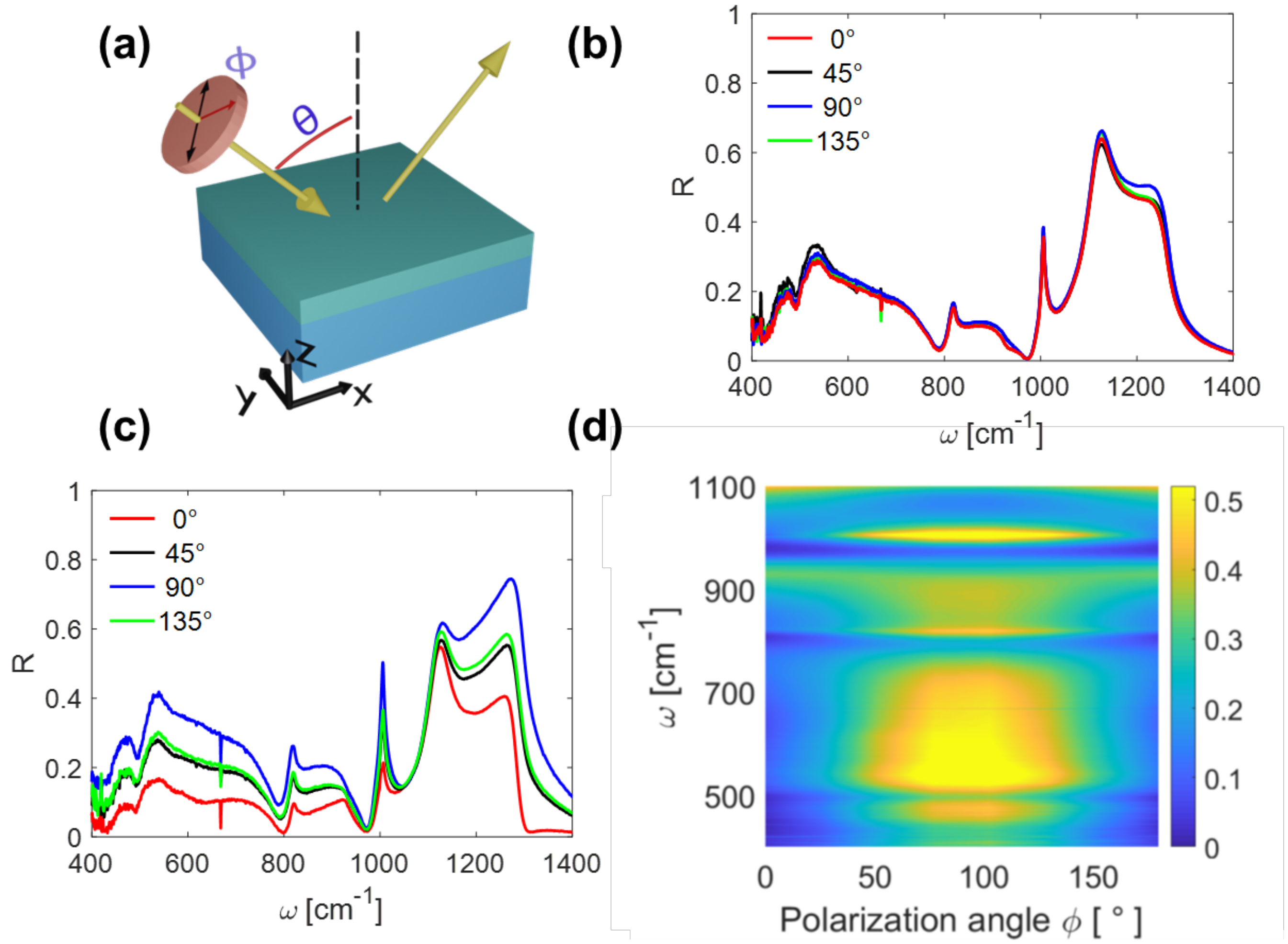

**Figure 2.** (a) Sketches of investigated experimental configuration; polarization-dependent (0°=p-pol, 90°=s-pol) reflection FT-IR spectra measured at (b) 15° and (c) 45° incidence angle from a α-$MoO_3$ film, grown on fused silica substrate using pulsed laser deposition; (d) surface plot of FT-IR reflection signal as a function of frequency and different polarization states of the incoming beam, measured at 45° incidence angle.

An ABB Bomen FTLA 2000 FT-IR [36], was also used to measure the reflectance of the samples after twelve months of the sample growth. The results as presented in Figure S3 agree very well with each other.

We note that the three RBs are contiguous in frequency and their sequential overlaps give rise to interesting spectral features: a) a polarization-independent perfect absorption condition at 972 $cm^{-1}$ (measured reflectivity less than 1%); b) a polarization tunable narrow band reflection peak at 1006 $cm^{-1}$. The behaviors of these two features are completely different when we consider 45° incidence angle. Results are displayed in Figure (2c). We note that the perfect absorption condition is almost preserved for both p- and s-polarizations; at 45° the experimental minimum reflectivity for both s- and p-polarization ranges between 1% and 2%. On the other hand, a strong modulation of $MoO_3$ film infrared spectral features with the polarization of the incoming light (Figure 2d) has been experimentally observed at 1006 $cm^{-1}$. Rotating the polarization state of incoming light (as highlighted in the legend) modifies both resonance intensity and width. It is worth noting that the polarization-dependent reflection peak at $\omega_{max}$=1006 $cm^{-1}$ with a full width at half maximum (FWHM) $\Delta\omega$=17 $cm^{-1}$ corresponding to a quality factor as high as $Q=\omega_{max}/\Delta\omega$ ~60 is obtained in a lithography-free polar film. We note that the reflection peak is not a pure Lorentzian resonance. In order to provide a more accurate evaluation of the resonance linewidth we considered two Lorentzian-shaped curves respectively fitting the inner and the outer part of the experimental data. Results are reported in the supporting material (Figure S4): the FWHM of the experimental resonance has been then retrieved by taking the average between the FWHMs of the Lorentzian curves and the maximum semi-dispersion as FWHM± Δ(FWHM) = (19 ±3) $cm^{-1}$. Thus the Q factor has been evaluated as Q±ΔQ=53±8. We finally include in the supporting material ( Figure S5) the reflection spectra of the previously mentioned monoclinic $\beta$–$MoO_3$ film (XRD pattern depicted in Figure S1) for an incidence angle of 45° and several polarization angles. We note that the high-Q polarization-dependent reflection peak at 1006 $cm^{-1}$ does not appear. Indeed this feature is specifically related to the $\alpha$-$MoO_3$ optical phonon along the crystal z-axis ($OPh_z$) [18].

### 3.2 Theoretical models.

Theoretical modeling of the optical properties of the polycrystalline $\alpha$-$MoO_3$ film are performed considering the following observations. (1) The AFM images of surface morphology reported in Figures (1c) and (1d) show that the grain sizes are much smaller than the infrared wavelengths in the measurements. (2) The almost negligible in-plane anisotropy demonstrated by the reflectance measurements at 15° angle of incidence (Figure 2b) implies that the crystallite grains in the material are nearly randomly oriented. (3) The thickness of $\alpha$-$MoO_3$ film varies from about 2000 nm to 2300 nm, as provided by the profilometer (see Figure S2 of the supporting material file). Thus the material should be considered isotropic and homogenous, and an average thickness of 2200 nm is used in the simulation for this sample.

Conventional dispersion analysis of homogeneous composites seeks the effective dielectric function based on the individual constituents, also known as the effective medium theory (EMT) [37]. For example, polycrystal materials with various crystallite sizes can be predicted using a modified EMT proposed by Mayerhöfer [38,39].

We also note that phonon frequencies in the polycristalline sample can be shifted away from the TO and toward the LO position when compared with a perfect crystal. For instance, the resonance at 1006 $cm^{-1}$ in Figure 2b and 2c is shifted with respect to the bulk $\alpha$-$MoO_3$ $\omega_{z,TO}$ = 957 $cm^{-1}$ [18]. This may be caused by the random orientation of crystallites in polycrystalline material, or the effect of air inclusions with an unknown volume fraction since a rough surface of $\alpha$-$MoO_3$ film is observed. Due to such unknown parameters, EMT such as the Maxwell-Garnett theory or an arithmetic average of the principle dielectric functions did not provide a satisfactory agreement with the measured spectrum.

Therefore, an isotropic Lorentz model with three oscillators, which roughly correspond to the frequency values of the oscillators in the *x*-, *y*-, and *z*-directions of a perfect crystal $\alpha$-$MoO_3$, is used to model the effective dispersion of the film:

$$\varepsilon(\omega) = \varepsilon_{inf} + \sum_{i=1}^{3} \frac{S_i \omega_i^2}{\omega_i^2 - i\gamma_i\omega - \omega^2}; \qquad (1)$$

The resonance frequencies $\omega_i$, oscillator strengths $S_i$, damping coefficients $\gamma_i$ in Eq. (1), with $i$ = 1,2,3, and $\varepsilon_{inf}$, are determined as fitting parameters by minimizing the RMS deviation between the calculated and measured reflection spectra. The fused silica substrate is modeled using the optical constants from Ref. [40].

Initially, we used [41] to calculate the reflectance for anisotropic stratified media. However, since the $\alpha$-$MoO_3$ film behaves as an isotropic medium we used the standard transfer matrix method for multilayer structures [37] in order to improve the speed and the efficiency of the fitting algorithm. Results obtained with the two methods are in perfect agreement. A thickness of 2200 nm is used in the calculation. The described fitting procedure applied to the experimental reflection spectra at 15° and 45° angle of incidences, allowed us to retrieve the parameters for the polycrystalline film with a RMS deviation of 0.054. The obtained parameters are listed in Table 1 with a typical error bound of 20% considering uncertainties in the measurements and fitting.

**Table1:** Fitted Lorentz oscillator parameters for the polycristalline $\alpha$-$MoO_3$ film.

| | $S_i$ | $\omega_i$ [$cm^{-1}$] | $\gamma_i$ [$cm^{-1}$] |
|---|---|---|---|
| 1 | 1.224 | 560 | 151 |
| 2 | 0.100 | 841 | 33.0 |
| 3 | 0.023 | 1005 | 3.74 |
| $\varepsilon_{inf} = 2.69$ | | | |

As mentioned, there exist a shift of the phonon frequencies towards LO for the resonators when compared with the values for a perfect crystal $\alpha$-$MoO_3$ reported in [18]: $\omega_{y,TO}$ = 545 $cm^{-1}$ , $\omega_{x,TO}$ = 821 $cm^{-1}$ and $\omega_{z,TO}$ = 957 $cm^{-1}$.

Figure 3 shows the comparison between the modeled and the measured reflectance spectra for s- and p-polarized incident fields at 45° angle of incidence. In Figure 4 we compare the measured reflectance spectra at 15° of incidence with the theoretical predictions obtained with the fitted parameters evaluated from the previous data. In both cases, the model calculation is in reasonable agreement with the experiment, except for 600 $cm^{-1}$ < $\omega$ < 1000 $cm^{-1}$. The fit of the sample deposited at 400 °C is not as good as that of the samples deposited at other temperatures. A better agreement could be obtained if more oscillators were used. However, this was not done due to the lack of information, and it is not the focus of this work. Overall, the RMS deviation between the model and experiments is within 0.06 for the entire spectrum. Details on the modeling of different samples and the effect of thickness are reported in Figures S6 and S7 of the supporting material, respectively.

As a final discussion, we focus on the small discrepancies between theoretical and experimental curves. It is worth mentioning that the reflectance measurement is performed with a measuring spot diameter of the order of a few millimeters. The inhomogeneity of the sampling area with varying surface roughness will result in the phonon frequency shifts to multiple positions. This effect leads to the deviation between experimental and theoretical data evaluated from a fixed frequency-oscillator model for calculation.

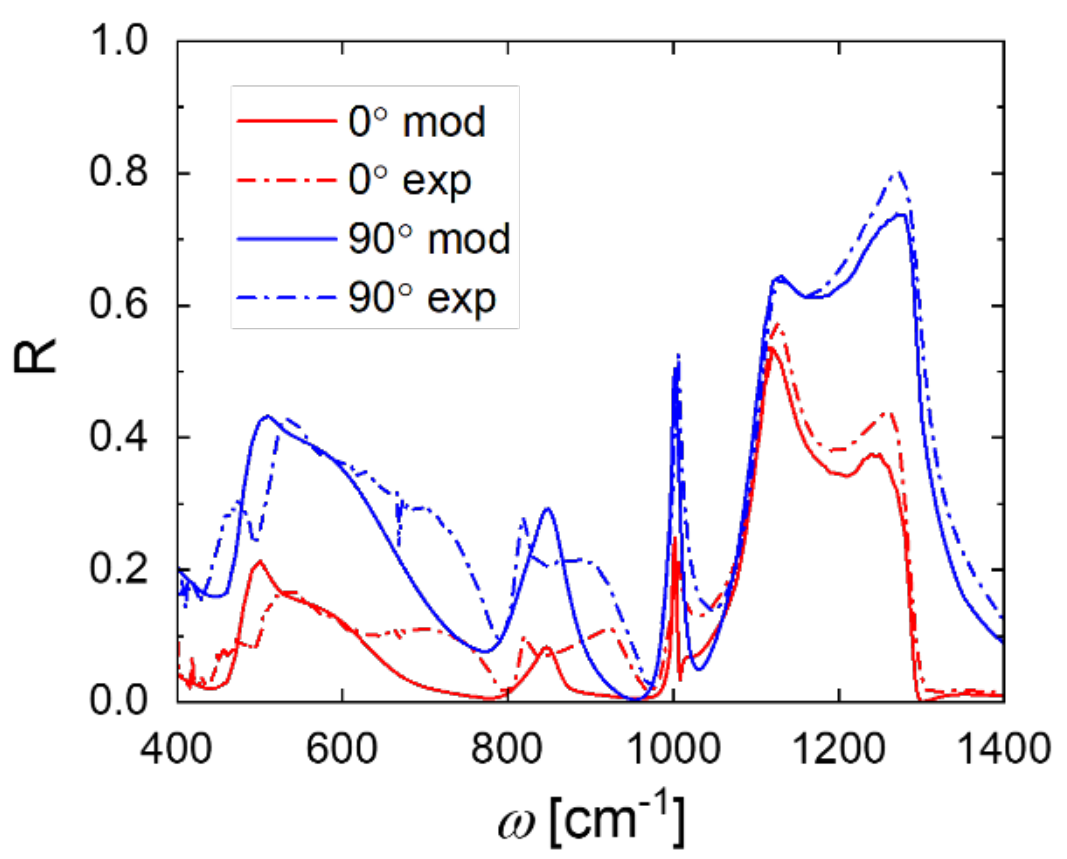

**Figure 3:** Comparison between the modeled (dash-dotted line) and the measured (solid line) reflectance spectra for s- and p-polarized incident fields at 45° angle of incidence.

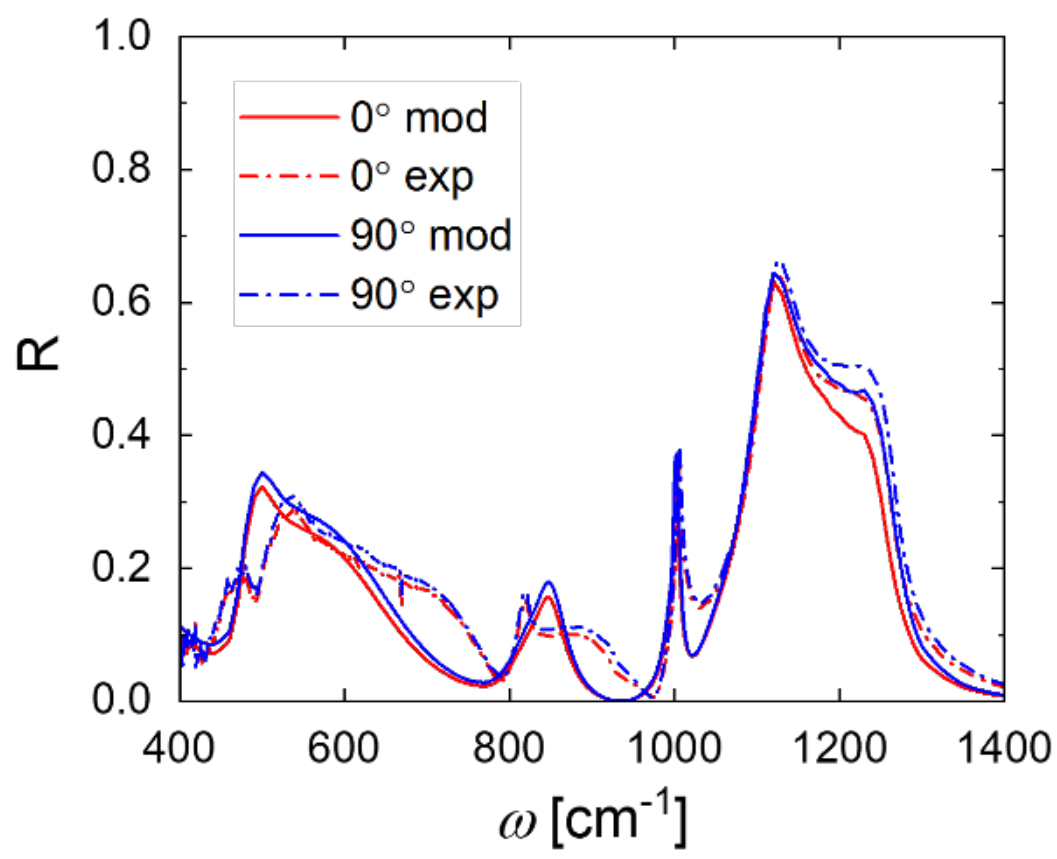

**Figure 4:** Comparison between the modeled (dash-dotted line) and the measured (solid line) reflectance spectra for s- and p-polarized incident fields at 15° angle of incidence.

## 4. Conclusions

Flat optics requires the design of optical components into thin, planar, films to be integrated into photonic platforms. The use of vdW materials, leads to 2D flat optics with ultra-compact and tunable devices. Nevertheless, atomically thin optical elements suffer from alignment issues and will low possibilities of far-field applications. To overcome this limitation and achieve control and tuning of the spectral features over a large area, we prepared and investigated $\alpha$-$MoO_3$ films using pulsed laser deposition. Although deposition parameter optimization is still required for the definition of a process to synthesize $MoO_3$ films with a high degree of crystallinity, our experimental findings show remarkable spectral features of the obtained polycrystalline films. Specifically, we reported both a polarization-independent perfect absorption behavior at 962 $cm^{-1}$ well preserved for a broad angular incidence range, starting from normal incidence and an enhanced tunability vs. light polarization angle of a narrow band reflection peak at 1006 $cm^{-1}$ with a Q factor of 53±8. The obtained high dynamic range of $\Delta R$=0.3 with off-normal-excitation (which can be improved with increased incidence angle, see Figure S8 in supporting material) is reliable and repeatable and results in wide tunability that may have great potential for label-free biochemical sensing application and narrow-band detection in the IR range without the use of time- and cost-consuming lithographic processes. In particular, the investigated sharp resonance can find applications to identify a spectral marker associated with specific moieties such as phenylalanine [42] tryptophan [43] to give some examples. We stress the point that the low fabrication cost and the Q-factor are not the only two relevant parameters to take into account. The possibility to operate with large sensing area without the need for microscopy or near-field techniques is an important benefit. Our large-area samples only require a basic far-field source/detector scheme, making them suitable for low-cost, mass distribution devices.

## Acknowledgements

K.A. acknowledges support from the Air Force Office of Scientific Research under Award Number FA9550-22-1-0300. K.A. and M.C.L. also acknowledge the support from University La Sapienza for the Visiting Professor Program 2020 (Bando Professori Visitatori 2020). M.C, M.C.L, M.A. and Z.M.Z. acknowledge the KITP program 'Emerging Regimes and Implications of Quantum and Thermal Fluctuational Electrodynamics' 2022, where part of this work has been done. This research was supported in part by the National Science. Foundation under Grant No. PHY-1748958.

C.Y. was supported by the National Science Foundation (CBET-2029892).

## Supporting Material

**Large-area polycrystalline α-MoO3 thin films for IR photonics**

Maria Cristina Larciprete[1], Daniele Ceneda[1], Chiyu Yang[2], Sina Abedini Dereshgi[3], Federico Vittorio Lupo[4], Maria Pia Casaletto[5], Roberto Macaluso[5], Mauro Antezza[6,7], Zhuomin M. Zhang[2], Marco Centini [1] and Koray Aydin [3]

[1] Dipartimento di Scienze di Base ed Applicate per l'Ingegneria, Sapienza Università di Roma, Rome, 00161 Italy

[2] George W. Woodruff School of Mechanical Engineering, Georgia Institute of Technology, Atlanta, GA 30332, USA

[3]Department of Electrical and Computer Engineering, Northwestern University, Evanston, Illinois 60208, USA

[4] Department of Engineering, University of Palermo, Palermo, 90128, Italy

[5] National Research Council (CNR), Institute for Nanostructured Materials (ISMN), Palermo, 90146, Italy

[6] Laboratoire Charles Coulomb (L2C), UMR 5221 CNRS-Université de Montpellier, F- 34095 Montpellier, France

[7] Institut Universitaire de France, 1 rue Descartes, F-75231 Paris Cedex 05, France

(*) Corresponding author:

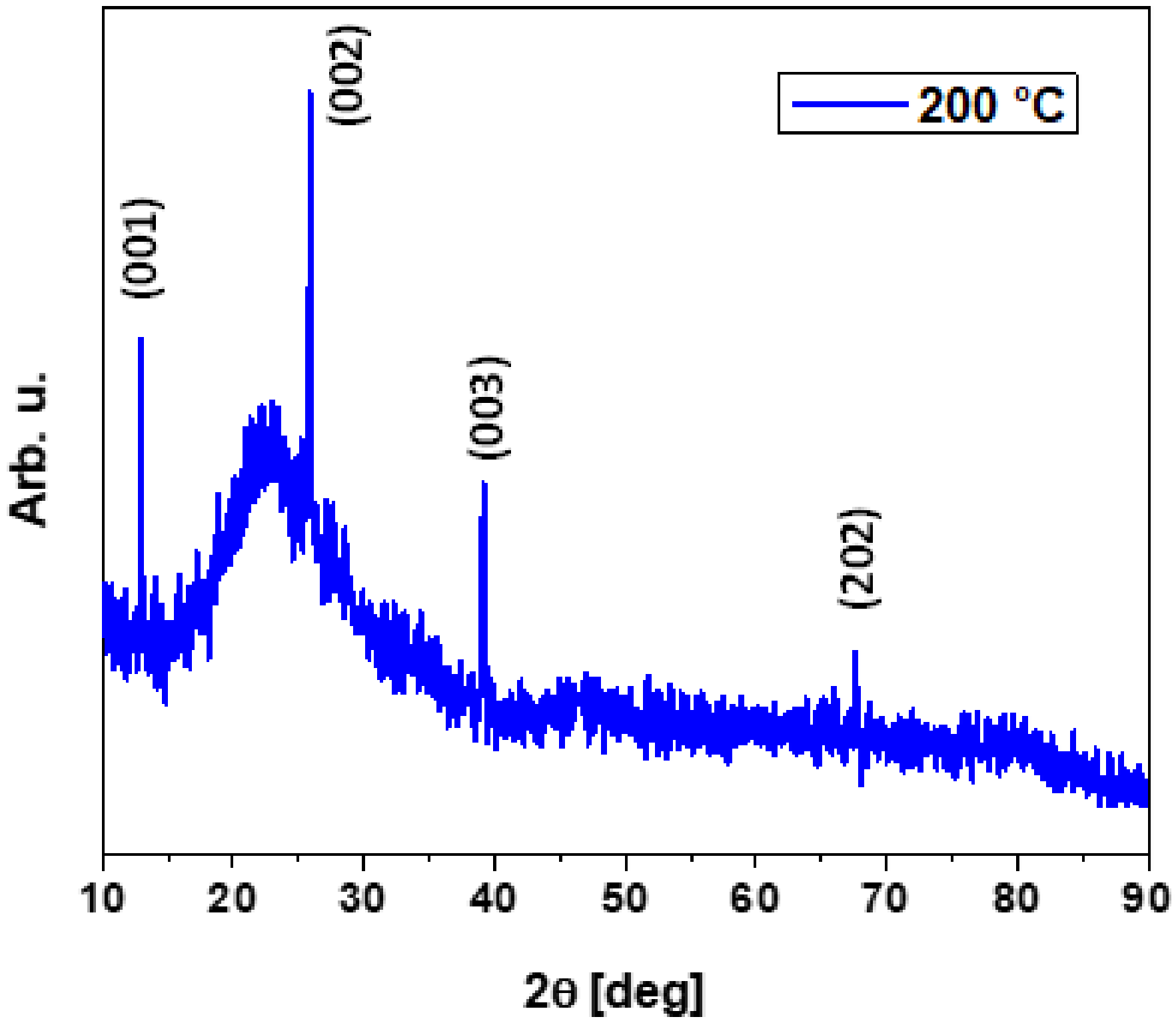

**Figure S1.** X-Ray diffraction (XRD) pattern of a $MoO_3$ film deposited at 200 °C and 0.1 mbar oxygen pressure. The assigned peaks correspond to the monoclinic β–phase of $MoO_3$ (ICDD 00-047-1320 card).

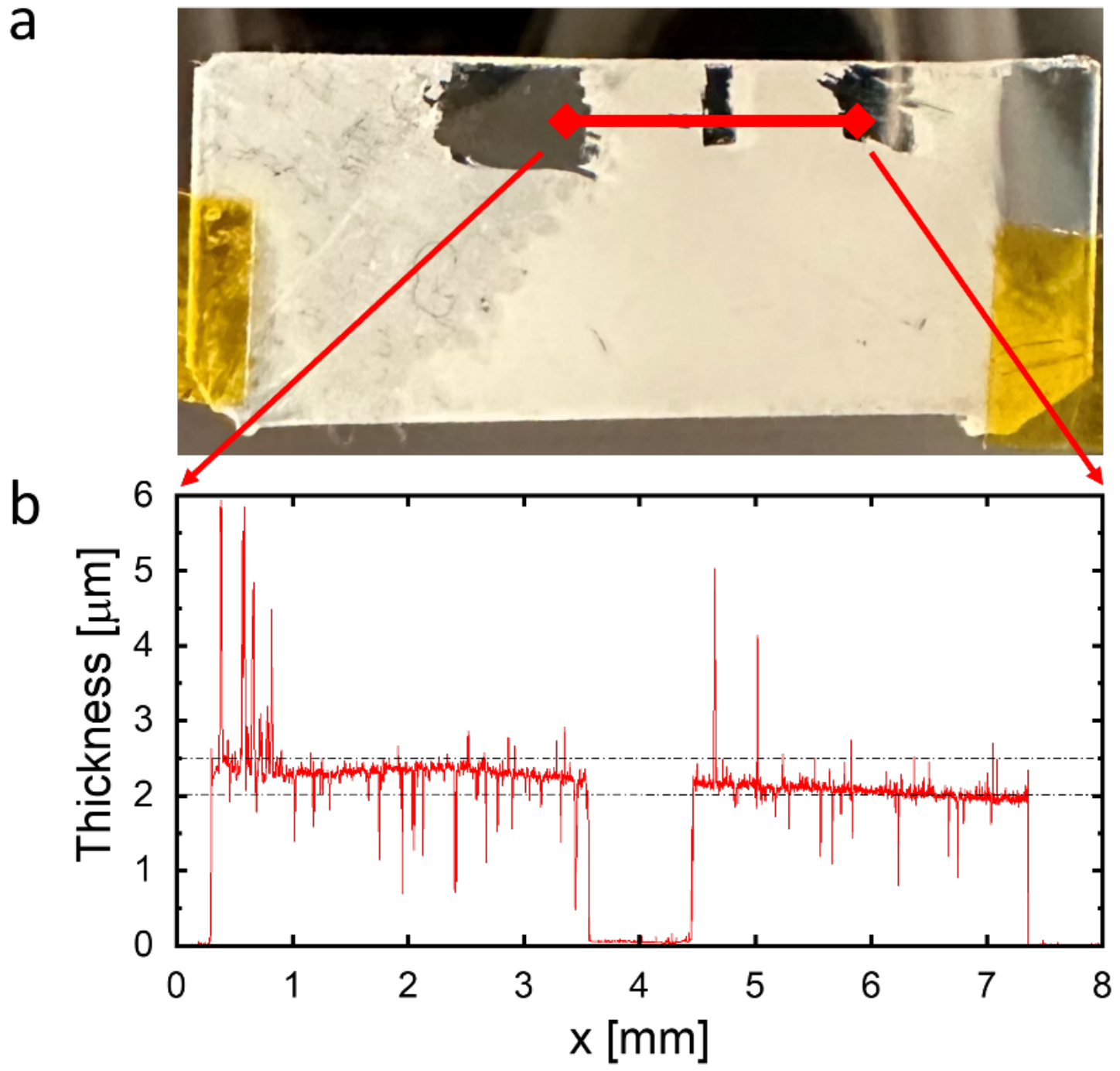

**Figure S2:** (a) Photo of the scratched $MoO_3$ film deposited at 400 °C, and (b) the thickness profile measured by Dektak 150 profilometer. The average film thickness is approximately 2200 nm with a root-mean-square (RMS) of 180 nm.

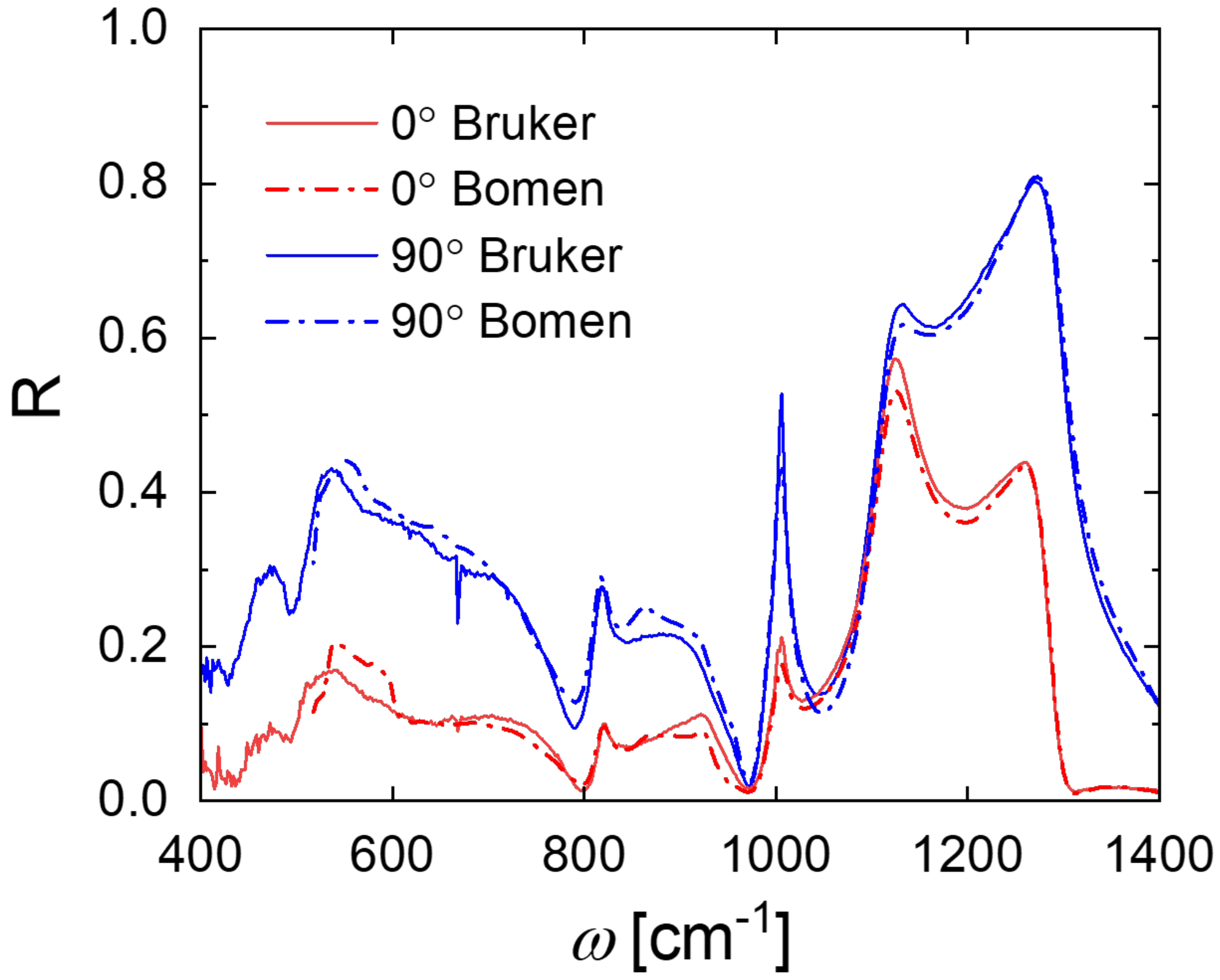

**Figure S3.** Comparison of the reflectance spectra measured by two FT-IR instruments, Bruker (solid line) and Bomen (dash-dotted line), at 45° incidence angle for the sample deposited at 400°C. The Bomen data were taken after twelve months of the film's growth. The RMS deviation between the two measurements is less than 0.02, which is within the measurement uncertainties, suggesting that the alpha $MoO_3$ films are stable in ambient conditions.

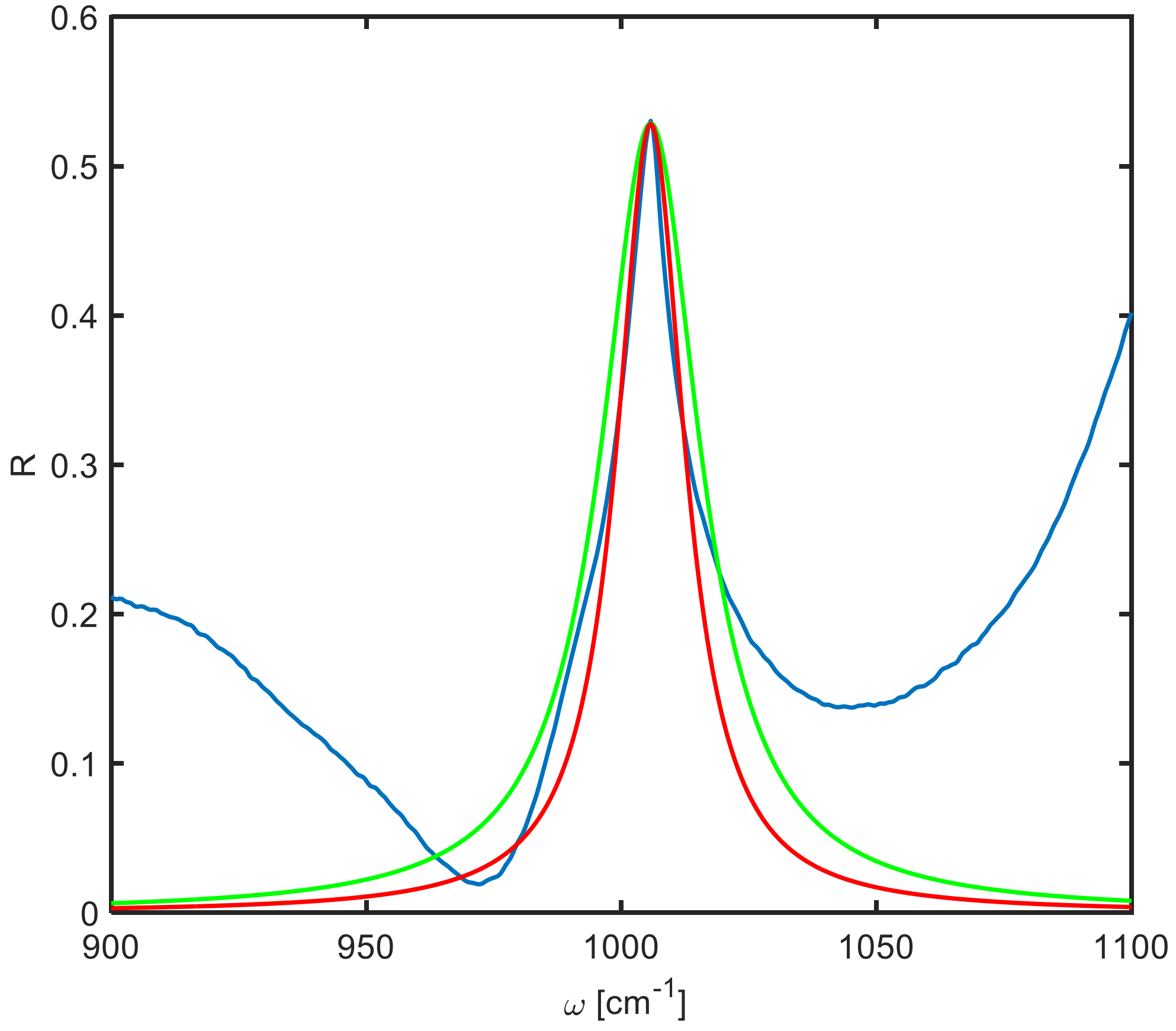

**Figure S4:** Zoom of the FT-IR reflection spectrum depicted in Figure 2c for s-polarized incident field (blue line). The resonance linewidth has been evaluated by considering two Lorentzian shaped curves to take into account the upper and the lower part of the resonance. In this way the FWHM of the experimental curve has been retrieved: FWHM ±Δ(FWHM)= 19±3 $cm^{-1}$.

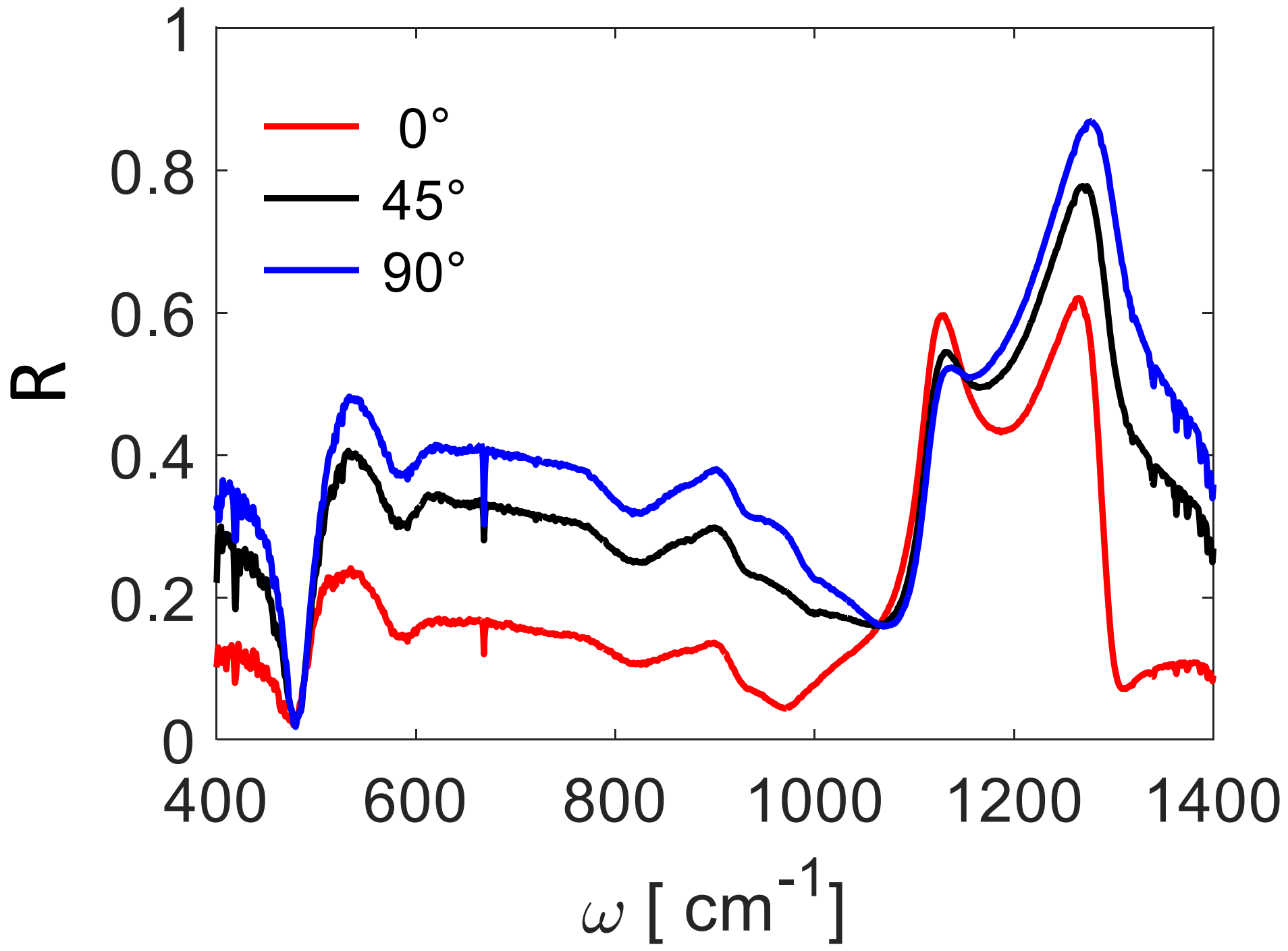

**Figure S5.** Measurements of spectral reflectance of monoclinic $MoO_3$ film deposited at 200°C. Incidence angle was set to 45° and several linear polarizer angles were employed (see legend, 0°= p-pol, 90°=s-pol).

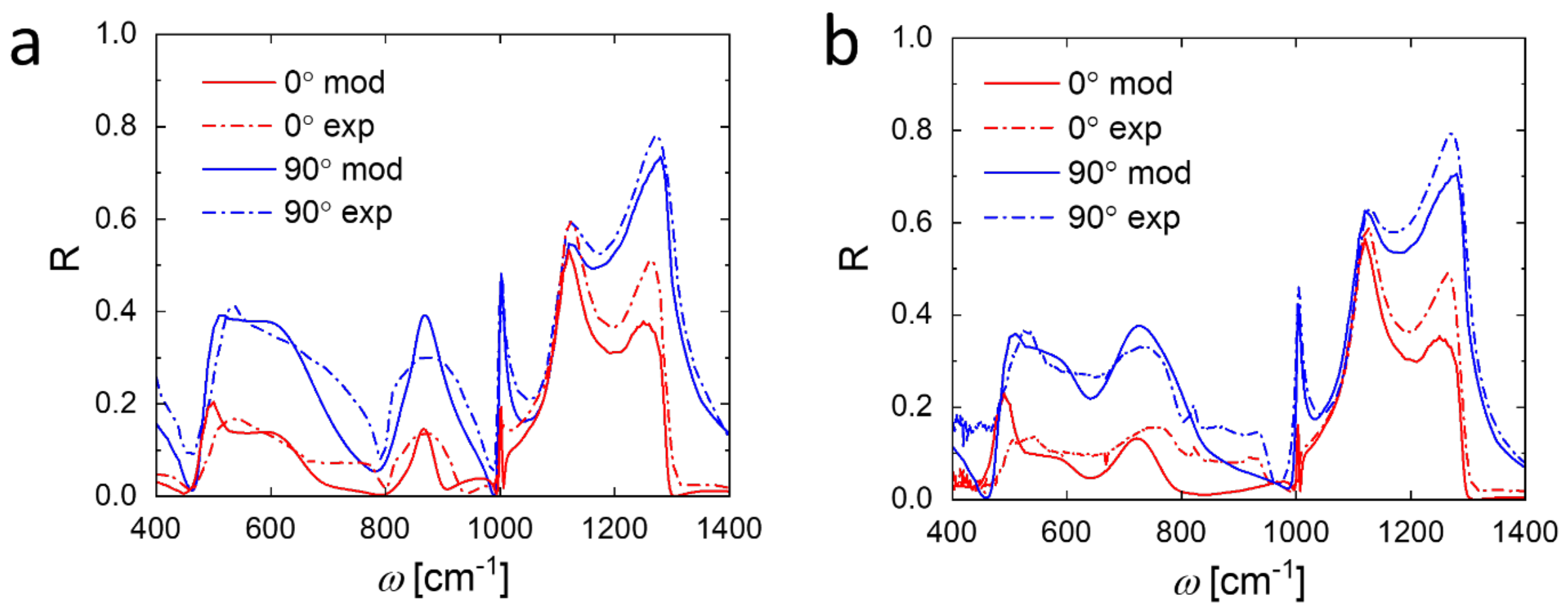

**Figure S6.** Measured (solid line) and calculated (dash-dotted line) FT-IR reflectance spectra at 45° incidence angle for the α-$MoO_3$ films deposited at (a) 300°C and (b) 500°C. The fitting has RMS of 0.049 (0°-polarized) and 0.070 (90°-polarized) for Figure (a) with parameters $t = 2100$ nm, $\varepsilon_{inf} = 2.16$, $S_1 = 0.782$, $\omega_1 = 590.0$ cm$^{-1}$, $\gamma_1 = 108.5$ cm$^{-1}$, $S_2 = 0.131$, $\omega_2 = 853.8$ cm$^{-1}$, $\gamma_2 = 32.88$ cm$^{-1}$, $S_3 = 0.012$, $\omega_3 = 999.0$ cm$^{-1}$, $\gamma_3 = 3.0$ cm$^{-1}$. The fitting has RMS of 0.051 (0°-polarized) and 0.055 (90°-polarized) for Figure (b) with parameters $t = 2400$ nm, $\varepsilon_{inf} = 1.86$, $S_1 = 0.346$, $\omega_1 = 590.0$ cm$^{-1}$, $\gamma_1 = 81.33$ cm$^{-1}$, $S_2 = 0.294$, $\omega_2 = 695.6$ cm$^{-1}$, $\gamma_2 = 74.56$ cm$^{-1}$, $S_3 = 0.010$, $\omega_3 = 1002$ cm$^{-1}$, $\gamma_3 = 3.0$ cm$^{-1}$.

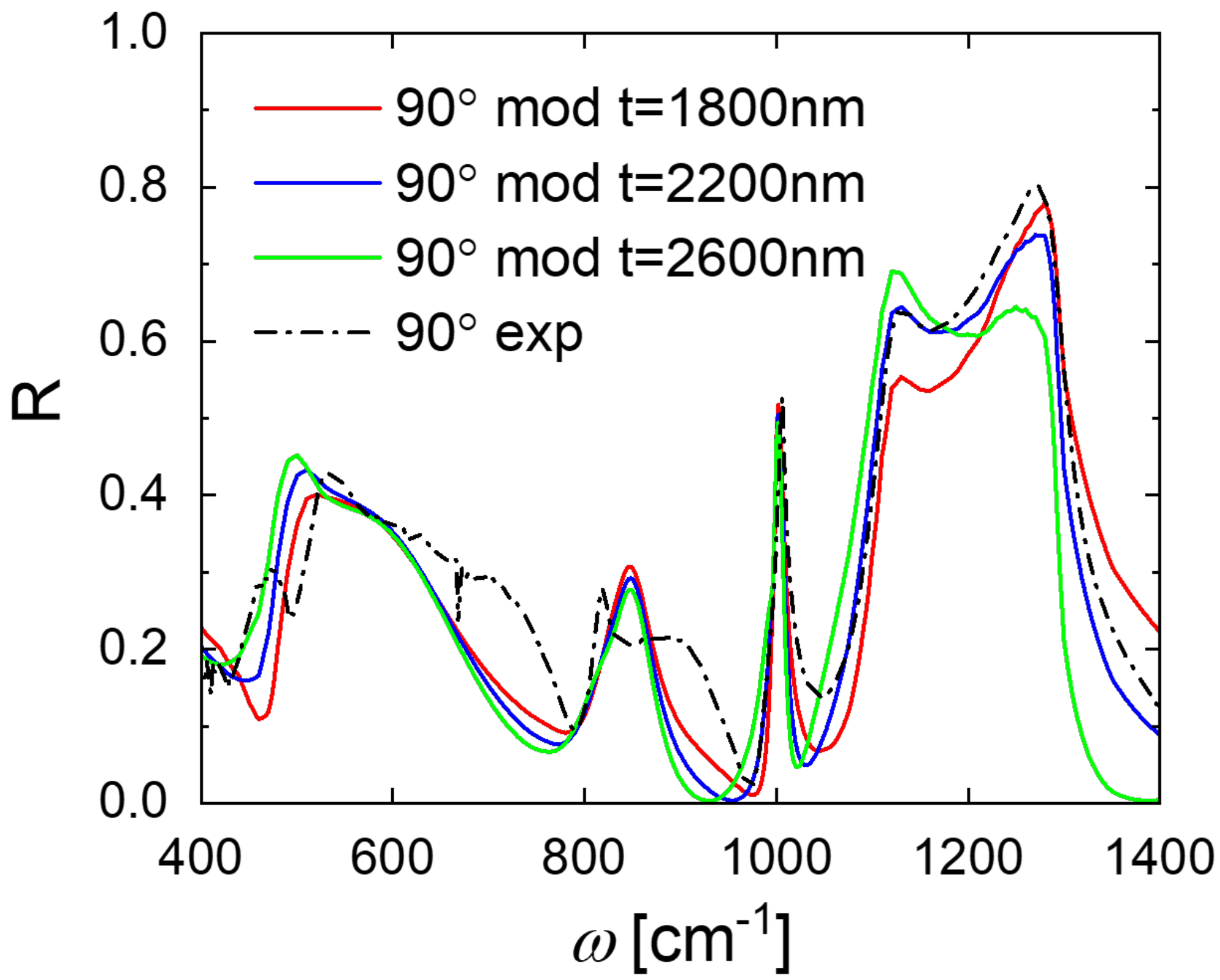

**Figure S7.** Comparison of the modeled reflectance at 15° incidence angle with α-$MoO_3$ film thickness *t* = 1800 nm and RMS = 0.076 (red line), *t* = 2200 nm and RMS = 0.074 (blue line), and *t* = 2600 nm and RMS = 0.119 (green line) for the film deposited at 400°C. In general, a similar RMS can be modeled for α-$MoO_3$ film thickness varying between 1800 nm to 2200 nm.

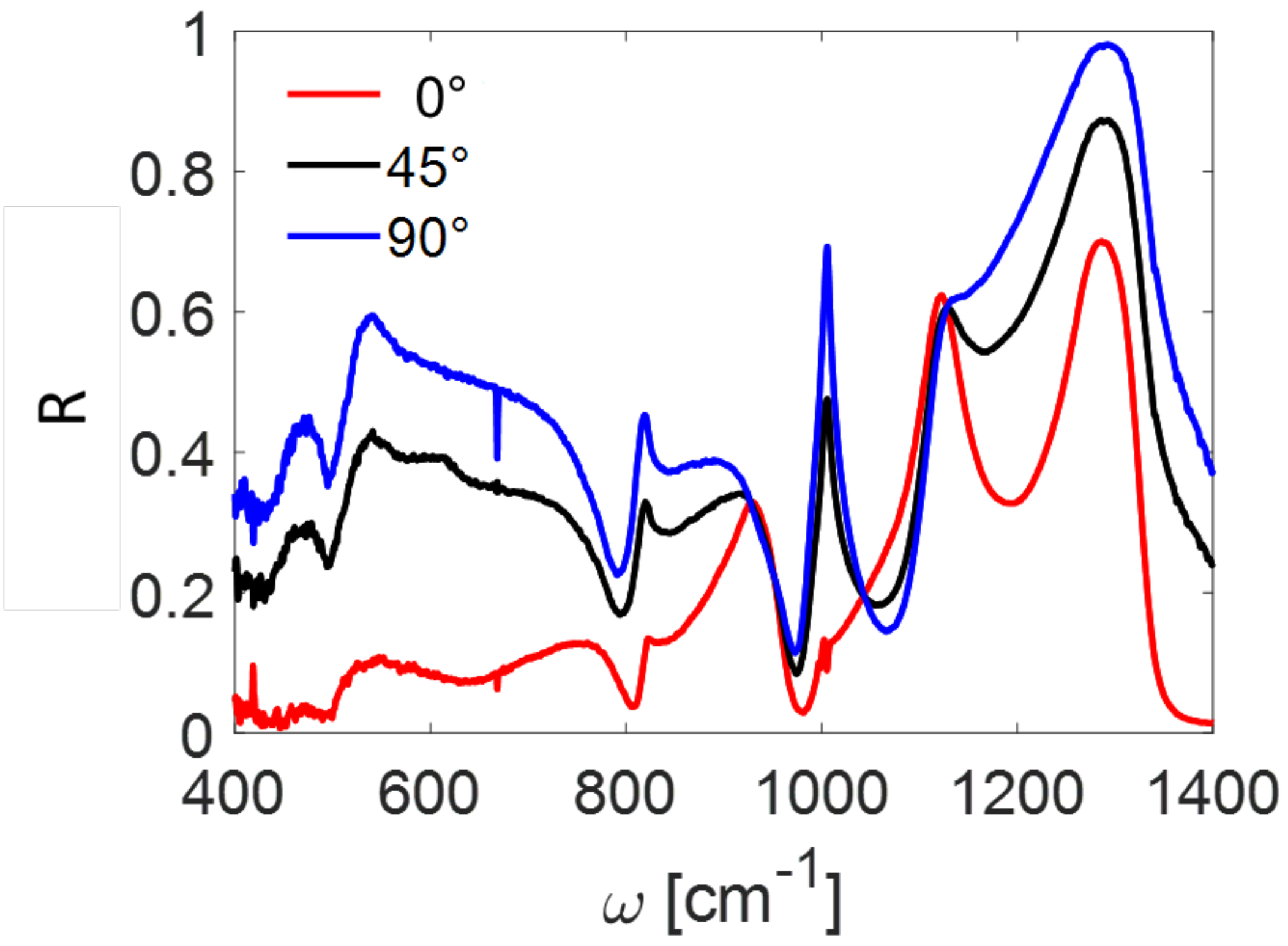

**Figure S8.** FT-IR reflectance spectra measured at 60° incidence angle for the sample deposited at 400°C as a function of different polarization state of the incoming light (see legend, 0°= p-pol, 90°=s-pol). The obtained dynamic range is ΔR = 0.58.